\documentclass[sigconf]{acmart}

\renewcommand\footnotetextcopyrightpermission[1]{} 

\setcopyright{none}

\acmConference[SBES 2026]{40th Brazilian Symposium on Software Engineering}{September 8–11, 2026}{São Paulo, SP, Brazil}

\AtBeginDocument{
    
}

\begin{document}

\title{Design-System-Aware Development with AI: Evaluating Productivity and Design Consistency}

\author{Luciane Silva}
\affiliation{%
  \institution{CI\&T \& UFU}
  \city{Uberlândia, MG}
  \country{Brazil}
  }
\email{lucianefs@ufu.br}

\author{Thayssa Rocha}
\affiliation{%
  \institution{Zup Innovation \& UFPA}
  \city{Belém, PA}
  \country{Brazil}
}
\email{thayssa.rocha@zup.com.br}

\author{Nicole Davila}
\affiliation{%
  \institution{Zup IT Innovation \& UFRGS}
  \city{Novo Hamburgo, RS}
  \country{Brazil}
}
\email{nicole.davila@zup.com.br}

\author{Gustavo Pinto}
\affiliation{%
  \institution{UFPA}
  \city{Belém, PA}
  \country{Brazil}
}
\email{gpinto@ufpa.br}

\renewcommand{\shortauthors}{Silva et al.}
\renewcommand{\shorttitle}{Design-System-Aware Development with AI}


\begin{abstract} 
Design Systems (DS) help standardize front-end development, yet developers still face challenges when translating high-fidelity mockups into consistent, production-ready interfaces. Although AI-assisted tools have emerged as a potential solution, empirical evidence on their effectiveness within DS-centered workflows remains limited. This paper reports a controlled experiment conducted at a large Brazilian enterprise that compares manual development, DS-only development, and DS-aware AI-assisted development across Angular, iOS, and Android stacks. Results from two experimental cycles show that AI assistance significantly reduced time-to-delivery (by 46.7\% to 69.4\%), increased task completeness, and decreased performance variability. Analysis of break patterns further suggests reduced workflow friction and smoother task execution. These findings provide empirical evidence that DS-aware AI tools can significantly accelerate development, improve design fidelity, and yield practical benefits for industrial front-end workflows.

\end{abstract}

\keywords{Artificial Intelligence, Design System, Productivity, Front-end Development, Software Engineering}

\frenchspacing
\maketitle

\section{Introduction}
Front-end development plays a central role in delivering digital products at scale. In large enterprises, this development must frequently conform to strict standards. Design Systems (DS) mitigate inconsistencies by providing reusable components; however, productivity still varies depending on the developer's ability to interpret and implement organizational conventions \cite{lamine2022}.

Recent advances in Artificial Intelligence (AI) present new opportunities \cite{RSLLLMES}. Nevertheless, generic AI tools operate without the context of internal corporate conventions and enterprise DS definitions \cite{PintoEtAlCAIN2024}. Consequently, developers often receive suggestions that are syntactically correct but visually inconsistent and non-compliant with enterprise governance.

To address this gap, this study investigates the following \textbf{Research Question:} \textit{How does the use of AI combined with a Design System affect the time required and execution characteristics of front-end development tasks?} 

This paper makes three key contributions to the software engineering industry: 
(1) a controlled industrial experiment involving 49 professional developers across Angular, iOS, and Android stacks;
(2) empirical evidence on the impact of Design-System-aware AI assistance on development productivity, dispersion, and task completeness;
(3) practical insights and lessons learned for organizations adopting AI-assisted front-end development workflows.

\section{Methods}
This section describes the study design, participants, procedures, and evaluation metrics used in the two experimental cycles.

\subsection{Study Design and Participants}
The study was structured as a controlled \textit{between-subjects} experiment conducted remotely and asynchronously in two independent cycles (June and October 2025). The sample consisted of 49 front-end and mobile developers (spanning Angular, iOS, and Android stacks) from Zup Innovation, a large Brazilian technology corporation. 

To reduce potential biases related to platform-specific proficiency or professional maturity, participants were balanced by their technical stack and seniority level across three independent groups:
\begin{itemize}
    \item \textbf{Manual (7 participants):} Followed a traditional workflow without assistance from structured tools.
    \item \textbf{Design System (21 participants):} Applied standardized components, layout tokens, and patterns from the organization's DS.
    \item \textbf{AI-Assisted (21 participants):} Used an internal AI tool (StackSpot AI) contextualized with the enterprise DS.
\end{itemize}

\textit{Note:} The manual group was smaller because it was only evaluated during the first cycle to establish a baseline.

\subsection{Experimental Task and Procedures}
\begin{figure}[htbp]
  \centering
  \includegraphics[width=0.95\columnwidth]{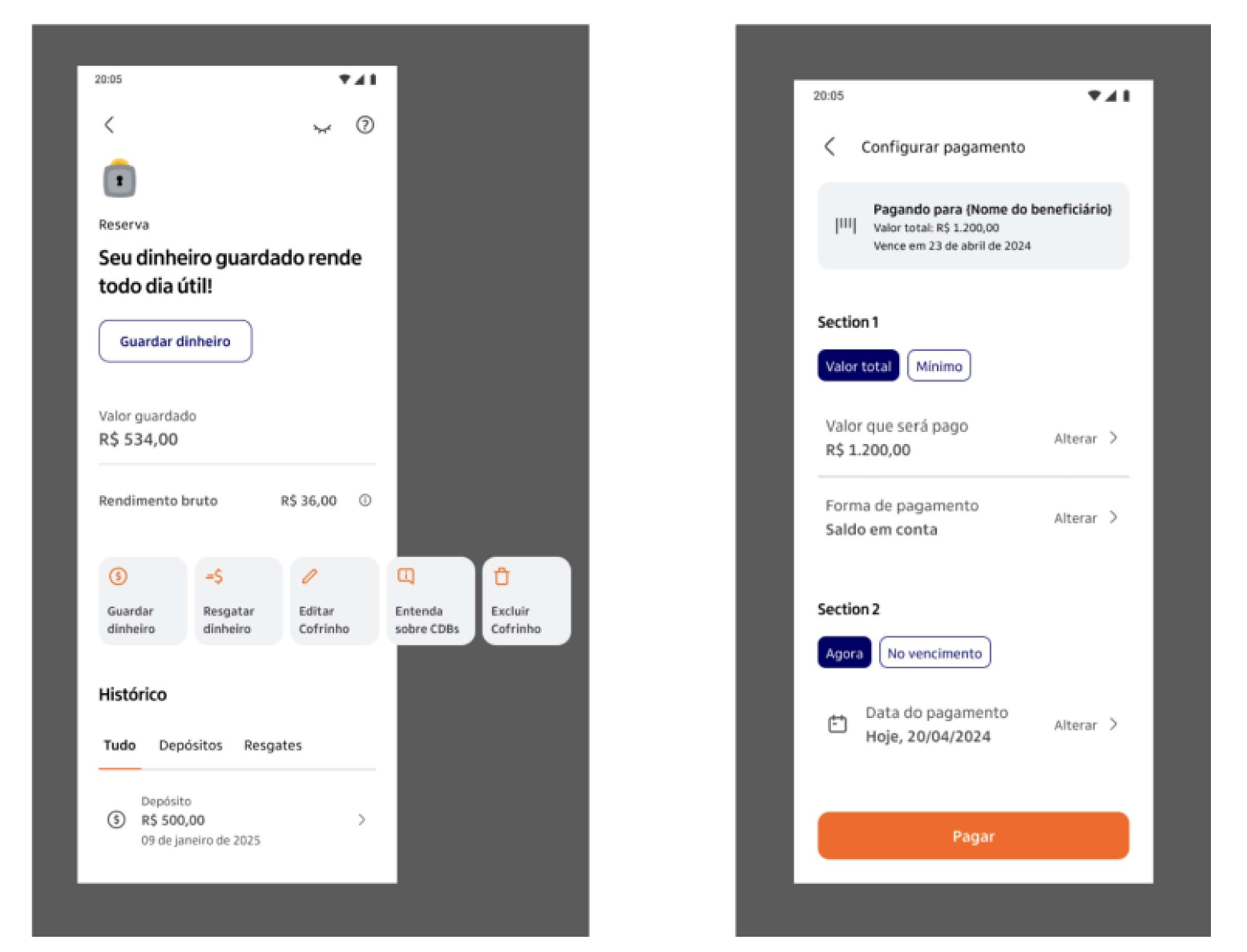}
  \caption{Example of a high-fidelity mockup used in the experiment.}
  \label{fig:mockup}
\end{figure}

All participants were tasked with implementing two screens based on a provided high-fidelity mockup (similar to Figure \ref{fig:mockup}) within a maximum duration of an 8-hour workday. They were expected to faithfully reproduce visual elements, component hierarchies, spacing rules, and responsiveness. 

The experiment followed an incremental submission protocol. Participants submitted each implemented section as they completed it. Specialists immediately reviewed the submissions in real-time. If the implementation was correct, the exact completion time was recorded; if inconsistencies were found, the section was returned for correction, and the timer continued. This ensured precise measurement and consistent checkpoints. Notably, the first experimental cycle evaluated all three groups, while the second cycle retained only the DS and AI-assisted groups, as the manual baseline was already sufficiently established in the first round.

\subsection{Evaluation Metrics}
The experimental outcomes were evaluated based on two primary quantitative metrics:
\begin{itemize}
    \item \textbf{Productivity (Time-to-Delivery):} The total time required to complete the implementation, including active development time and natural work pauses (such as breaks).
    \item \textbf{Task Completeness:} Assessed via visual fidelity and the presence of essential mockup elements, verified by the reviewing specialists.
\end{itemize}

Additionally, qualitative data regarding break patterns and tool interaction were collected by the supervising specialists. These observations served as an indirect indicator of developers' cognitive load and workflow continuity throughout the task.
\section{Results}

This section presents the empirical findings of the controlled experiment addressing the central research question of this study: to what extent AI assistance, when aligned with an enterprise Design System, improves development performance in front-end and mobile workflows. The analyses consolidate both experimental rounds and include developers from Angular, iOS, and Android.

\subsection{Productivity, Dispersion, and Completeness}

Across all technological stacks, AI-assisted development substantially reduced the time required to implement the two high-fidelity interfaces. AI-assisted development significantly reduced completion time compared to both the manual baseline and DS-only development ($p < 0.05$). 

Furthermore, AI assistance drastically decreased performance variability among developers. As evidenced by the lower Standard Deviation (SD) in Table \ref{tab:productivity}, AI helps reduce extreme variations, producing more uniform and predictable results.

\begin{table*}[t]
\centering
\caption{Productivity (Time-to-Delivery) and Dispersion}
\label{tab:productivity}

\begin{tabular}{lcc|cc|cc}
\toprule
& \multicolumn{2}{c}{\textbf{Angular}}
& \multicolumn{2}{c}{\textbf{iOS}}
& \multicolumn{2}{c}{\textbf{Android}} \\

\textbf{Group}
& \textbf{Mean (min)}
& \textbf{SD}
& \textbf{Mean (min)}
& \textbf{SD}
& \textbf{Mean (min)}
& \textbf{SD} \\
\midrule

Manual
& 536.15 & 145.0
& 593.00 & 160.0
& 387.75 & 110.0 \\

Design System
& 216.75 & 85.0
& 374.00 & 112.0
& 213.25 & 75.0 \\

AI-Assisted
& 164.00 & 42.0
& 316.00 & 51.0
& 163.25 & 35.0 \\

\bottomrule
\end{tabular}
\end{table*}

Beyond speed, AI assistance yielded higher consistency. As shown in Table \ref{tab:completeness}, developers in the AI group achieved a higher average completeness rate regarding the required design elements and structural fidelity, supporting our initial claims.

\begin{table}[h]
\centering
\caption{Average Task Completeness by Group}
\label{tab:completeness}
\begin{tabular}{l c}
\toprule
\textbf{Group} & \textbf{Avg Completeness} \\
\midrule
Manual        & 68\% \\
Design System & 85\% \\
AI-Assisted   & 96\% \\
\bottomrule
\end{tabular}
\end{table}

Across all technological stacks, AI-assisted development substantially reduced the time required to implement the two high-fidelity interfaces. Table~\ref{tab:productivity} summarizes the consolidated results. Angular exhibited the highest manual baseline and the largest absolute reduction, with average completion time decreasing from 536.15 to 164.00 minutes (69.4\%). With iOS, participant performance showed the widest dispersion. AI support still reduced the completion time from 593.00 to 316.00 minutes (46.7\%). Android, which presented the lowest manual baseline, also benefited significantly from AI, with average time decreasing from 387.75 to 163.25 minutes (57.9\%).

Taken together, these findings show that AI assistance not only accelerates development but does so consistently across heterogeneous environments, independent of baseline complexity or platform-specific characteristics.

\begin{table}[t]
\centering
\caption{Consolidated Productivity Results}
\label{tab:productivity}
\begin{tabular}{l c c c}
\toprule
\textbf{Metric} & \textbf{Angular} & \textbf{iOS} & \textbf{Android} \\
\midrule
Manual (avg min) & 536.15 & 593.00 & 387.75 \\
Design System (avg min) & 216.75 & 374.00 & 213.25 \\
AI-Assisted (avg min) & 164.00 & 316.00 & 163.25 \\
AI vs. Manual & 69.4\% & 46.7\% & 57.9\% \\
AI vs. DS & 24.4\% & 15.5\% & 23.4\% \\
\bottomrule
\end{tabular}
\end{table}


\subsection{Cross-Stack Comparative Analysis}

Across the three technological stacks, we observed distinct patterns in how AI assistance influenced development performance. Angular exhibited the largest absolute gains, driven by a significantly higher baseline in the manual workflow. Developers without assistance frequently required extensive time to interpret layout rules, component hierarchies, and visual structures, resulting in the highest mean completion times among all stacks. With AI support, however, Angular showed one of the steepest reductions, indicating that the stack particularly benefits from automated guidance in translating design specifications into code.

iOS presented the greatest variability across participants and groups. This variability reflects the heterogeneity of iOS development workflows, differences in familiarity with UIKit or SwiftUI conventions, and the presence of repeated navigation and component patterns that can be interpreted in multiple ways. Although the relative reduction achieved by AI was substantial, the wide dispersion suggests that iOS performance is influenced more heavily by individual developer experience, tooling expertise, and architectural preferences than the other stacks.

Android, in contrast, displayed the lowest baseline times in the manual and Design System conditions. This may be attributed to more predictable layout structures, consistent component patterns, or greater fluency among participants in this stack. Despite starting from a comparatively lower baseline, Android still demonstrated meaningful reductions with AI assistance, highlighting that the benefits of AI support extend even to stacks where developers already perform efficiently. The contraction of the upper bound of completion times further suggests that AI helps reduce extreme variations, producing more uniform and predictable results.

Together, these differences illustrate that while all stacks benefit from AI assistance, the magnitude and nature of the impact vary: Angular gains are driven by complexity reduction, iOS by variability smoothing, and Android by incremental efficiency improvements over an already optimized baseline.

\subsection{Breaks and Workflow Friction}
Beyond completion times, we also examined the duration and distribution of breaks taken by participants during the development task. Break patterns offer an indirect but meaningful indicator of the level of friction experienced while interpreting and implementing design artifacts. Across both datasets, the AI-assisted group exhibited consistently shorter and less frequent breaks.

Participants in the Manual workflow recorded the longest pauses, with a maximum of 195 minutes in the R3 dataset. These prolonged interruptions suggest "dead-ends" when interpreting layout rules without structured guidance. The Design System group showed intermediate results, still exhibiting extended break periods (often between 60 and 110 minutes), indicating that the DS alone does not fully mitigate interpretive challenges.

In contrast, the AI-assisted group demonstrated shorter and more homogeneous break durations (0 to 90 minutes). Overall, these break patterns suggest reduced workflow friction and smoother task execution. Furthermore, they may indicate lower cognitive effort, enabling developers to progress more fluidly and with fewer interruptions throughout the task.

\section{Discussion}

In this section, we provide further discussion on this study.

\subsection{Summary of the findings}

Across both experimental cycles (June and October 2025) and all technological stacks, AI assistance meaningfully improved development outcomes. It reduced time-to-delivery by 46–69\%, increased task completeness, minimized extreme variations, and supported more stable and consistent execution. Although the Design System alone improved efficiency relative to manual workflows, its impact was systematically smaller than that of AI. These findings demonstrate that Design-System-aware AI tools can significantly enhance delivery speed, design fidelity, and operational consistency in front-end development workflows.

\subsection{Lessons Learned}
Based on the experimental observations, we highlight the following practical recommendations and lessons learned for industrial adoption:
\begin{itemize}
    \item \textbf{Contextualized AI produced better outcomes than generic assistance:} AI tools aware of the enterprise Design System avoided non-compliant code suggestions.
    \item \textbf{Design Systems alone improved productivity but did not eliminate interpretation effort:} Developers still struggle to combine complex components without AI guidance.
    \item \textbf{AI benefits varied by technological stack:} Engineering leaders should target AI adoption and optimize expectations based on the specific needs of each stack (e.g., complexity reduction in Angular vs. variability smoothing in iOS).
    \item \textbf{Organizational knowledge embedded in AI appears to be a key adoption factor:} Integrating the company's specific guidelines directly into the tool's context is essential for scaling consistent interfaces.
\end{itemize}

\subsection{Limitations}

This study has several limitations that should be considered when interpreting the findings. First, all participants were drawn from a single large Brazilian enterprise, which may limit the generalizability of the results to organizations with different cultures, workflows, or design governance structures. Second, the experimental task focused on implementing two predefined screens, a scenario that, while controlled, does not fully capture the broader complexity of real-world front-end development, including requirement refinement, integration with APIs, iterative review cycles, or long-term maintenance. Third, completeness and visual fidelity were evaluated by specialists rather than through automated or multi-reviewer assessments, introducing potential subjectivity in borderline cases. Finally, the AI assistance used in the experiment was specifically tailored to the organization's Design System. As such, the observed benefits may differ when using generic AI tools that lack similar contextualization. Future studies should explore more diverse tasks, broader participant samples, and longitudinal effects to provide a more comprehensive understanding of AI support in front-end workflows.

\subsection{Implications}

These findings carry direct implications for the software industry. Primarily, they provide empirical support for the effectiveness of Design-System-aware AI strategies: we demonstrated that AI tools cognizant of the Design System can significantly enhance delivery speed, design fidelity, and operational consistency in front-end workflows. Organizations operating in regulated environments or managing extensive product portfolios can therefore use this evidence to justify the integration of AI tools adapted to their internal standards. Secondly, the results allow for Development Optimization by Stack, as AI showed varied benefits: offering complexity reduction for Angular, variability smoothing for iOS, and incremental efficiency improvements over an already optimized baseline for Android. Engineering teams can, thus, use these insights to target AI adoption and optimize the development experience based on the specific needs of each stack.

Although our results are robust across three technological stacks, further studies in different organizational domains and project scales would help establish broader generalizability.

\section{Related Work}

Recent advances in AI, particularly Large Language Models (LLMs), have increased interest in AI-assisted software development. Prior work reports productivity gains and changes in developer workflows when using AI-based coding assistants~\cite{AlamiErnst2025, DavilaEtAl2024,bird2023}. In industry, researchers identified specific contextual knowledge as a critical factor in the effective adoption of AI-based tools~\cite{PintoEtAlCAIN2024}. Nevertheless, as discussed in a recent literature review~\cite{RSLLLMES}, most empirical evaluations focus on small-scale programming activities or controlled academic settings, offering limited insight into complex, design-constrained, and enterprise-scale development workflows.

In another direction, Design Systems (DS) have emerged as a foundational practice to improve consistency, reuse, and scalability in front-end development. The effectiveness of DS in practice depends not only on tooling and documentation but also on developers’ understanding, coordination mechanisms, and organizational context~\cite{lamine2022}. From a cognitive perspective, reducing extraneous cognitive load can improve task performance by allowing individuals to focus on essential problem-solving activities~\cite{sweller2019}. In the context of front-end development, this suggests that tools that reduce the interpretive effort required to translate design artifacts into code may improve productivity and workflow continuity.

Despite these advances, a clear gap remains at the intersection of AI-assisted development and DS-centered workflows. Few studies have empirically evaluated AI tools that are explicitly aware of and aligned with enterprise DS through controlled experiments conducted in real industrial environments. Moreover, the combined effects of AI assistance and DS on productivity, task completeness, and cognitive load remain underexplored.

This study addresses this gap by presenting a controlled industrial experiment comparing manual development, DS-only development, and DS-aware AI-assisted development across multiple technology stacks. By grounding the evaluation in real-world front-end tasks and enterprise constraints, this work complements prior research on AI-assisted programming and contributes empirical evidence on how contextualized AI tools can enhance productivity, consistency, and cognitive efficiency in design-constrained software development.

\section{Conclusion}

This study investigated the impact of AI assistance aligned with an enterprise Design System (DS) on the productivity and consistency of front-end and mobile development. Through a controlled experiment conducted in a real corporate environment, comparing manual, Design System-only, and AI-assisted workflows across three technological stacks (Angular, iOS, and Android), our results provide substantial empirical evidence on the value of AI in design-constrained development. 

Our main findings indicate AI support demonstrated a substantial reduction in time-to-delivery across all platforms. Productivity gains ranged from 46.7\% (iOS) to 69.4\% (Angular), and 57.9\% (Android), indicating that AI assistance consistently accelerates development. Furthermore, AI helped minimize extreme variations in completion times, promoting more uniform and predictable results. From a human-centered perspective, the observed reduction in task interruptions suggests a shift toward more sustainable and less cognitively fragmented development workflows.


\section*{Acknowledgments}

We thank the reviewers for their helpful comments. For partially supporting this work, we would like to thank INES.IA (National Institute of Science and Technology for Software Engineering Based on and for Artificial Intelligence) www.ines.org.br, CNPq (grant 408817/2024-0, 314680/2026-8, 308623/2022-3). 

\bibliographystyle{ACM-Reference-Format}
\bibliography{sample-base}

\end{document}